# The V/$V_m$ test for a sample of compact steep-spectrum sources
(August 1983)

V. K. Kapahi & D. G. Banhatti
Radio Astronomy Centre (TIFR), P. O. Box 1234,
Bangalore 560 012, India

**Summary / Abstract.** We apply the V/$V_m$ test to a subsample of compact steep-spectrum sources from a complete sample of radio sources selected at 2.7 GHz. We find that the <V/$V_m$> has a value intermediate between those found for samples of extended steep-spectrum sources and those of compact flat-spectrum sources. If the sample is split into two further classes of sources having more steep and less steep spectra, the <V/$V_m$> values for these then tally roughly with those found for the extended steep-spectrum and compact flat-spectrum classes of sources. Implications of this result are discussed.

## Introduction
The earliest surveys of radio sources showed that the extended sources had a steep spectrum and the unresolved compact sources had a flat (or inverted or complex) spectrum. It has recently been realized that there are also compact sources having a steep spectrum. These show up prominently in surveys at frequencies $\gtrsim$ 1 GHz. The evolutionary properties of various samples of extended steep-spectrum (ESS) and compact flat-spectrum (CFS) sources have been extensively examined through applying the $V/V_m$ (i.e., luminosity-volume) test to them, among other methods like the source count test. It has been found that the values of $<V/V_m>$ for these two classes are $\approx 0.74$ and $\approx 0.60$ respectively (see, e.g., Wills & Lynds 1978) for uniform relativistic world-models. If there is no evolution of the source population with cosmic epoch in a world-model, $<V/V_m>$ is expected to be ½.. A larger value indicates that the sources in the population under consideration were either more numerous and/or more luminous in the past. So these $<V/V_m>$ values show that the CFS sources evolve cosmologically and that the ESS sources evolve even more strongly. The evolutionary properties of compact steep-spectrum (CSS) sources have not been examined partly because they have been recognized as a separate class only recently and also because a large enough sample of such sources has only recently become available (Peacock & Wall 1981; 1982).

**The Sample**

We have selected the 34 unresolved (with a resolution of 2 to 4 arcsec) sources with steep spectra (spectral index $-d \log S / d \log \nu \equiv \alpha > 0.5$) from Peacock & Wall's (1981) 2.7 GHz sample complete down to 1.5 Jy. Thirteen of the sources do not have a measured redshift, but 5 of them have a magnitude estimated for the optically identified galaxy or quasar. We have estimated the redshifts of these sources from the straight line fits (Lang et al 1975) to the m - log z relation for radio galaxies and quasars. The remaining 8 sources are either empty fields to the limit of the PSS prints or have faint objects associated with them. We have calculated two sets of $V/V_m$ values for these, taking the redshifts to be 0.5 for one set and 1.0 for the other.

**Results**

The $V/V_m$ value for each source was calculated using the spectral index α between 2.7 GHz and 5 GHz (Peacock & Wall 1981) in the Einstein-de Sitter world-model. The mean value for all the 34 sources is intermediate between the values ≈ 0.60 for CFS sources and ≈ 0.74 for ESS sources, as found from several studies (see, e.g., Wills & Lynds 1978). If we split the sample into the 16 sources with α ≥ 0.7 and the remaining 18 with α < 0.7 we find that the $<V/V_m>$ values for these two classes are similar to those for ESS and CFS sources respectively. The various $<V/V_m>$ values are tabulated in Table 1. These values, however, are based on small samples and need to be confirmed by formulating larger independent samples of CSS sources.

**Table 1.** Values of $<V/V_m>$

| Sample | No. of sources# | $<V/V_m>$* |
|---|---|---|
| α ≥ 0.7 | 16 (5) | 0.69 +/- 0.069 |
|  |  | 0.71 +/- 0.067 |
| α < 0.7 | 18 (3) | 0.58 +/- 0.063 |
|  |  | 0.59 +/- 0.064 |
| All CSS | 34 (8) | 0.63 +/- 0.047 |
|  |  | 0.65 +/- 0.047 |

# The numbers in parentheses indicate the number of sources for which the redshift has been assumed.

* The upper line gives the value assuming a redshift of 0.5 for the unidentified and optically very faint objects. The lower line corresponds to an assumed redshift of 1.0 for these.

## Discussion

Our result that the value of $<V/V_m>$ for CSS sources is intermediate between those for CFS and ESS sources is consistent with Peacock & Walls's (1982) observation that the percentage of such sources in surveys at many frequencies between 408 MHz and 5000 MHz down to various flux density levels (the lowest being 0.015 Jy at 1400 MHz) is approximately the same. Since these surveys are an admixture of flat-spectrum and steep-spectrum sources, this indication of the CSS source count being roughly the same as their count is in accordance with the $<V/V_m>$ value we have found for the CSS sources.

It has been a general practice to divide extragalactic radio sources into the steep-spevtrum ($\alpha > 0.5$) and flat-spectrum ($\alpha < 0.5$) classes rather arbitrarily at the value $\alpha = 0.5$. No satisfactory physical explanation has been given of this division. Our result shows that, whatever be the physical explanation for any such spectral division, the demarcation at $\alpha = 0.5$ may not be meaningful in at least one respect, viz., cosmological evolution; and that a division at a higher value of $\alpha$ may be more meaningful.

To see if the division of CSS ($\alpha > 0.5$) sources into sources with more steep ($\alpha \geq 0.7$) and less steep ($\alpha < 0.7$) spectra shows up in other properties of the sources, we have calculated the 2.7 GHz luminosities of these sources (in the Einstein-de Sitter world-model with $H_0 = 50$ km.s$^{-1}$.Mpc$^{-1}$). If the higher $<V/V_m>$ value for sources with steeper spectra is because these are more luminous than the other class in accordance with the log P – $\alpha$ correlation found for steep-spectrum sources (MacLeod & Doherty 1972), this should show in the mean luminosity of these classes. We find that there is no significant difference between the mean values of log P for the two classes. We also note that the identification content (proportion of galaxies, quasars and empty fields) is not significantly different for the two classes. Thus, though the samples are rather small, we conclude that the difference in the evolutionary behaviour of these two classes of radio sources cannot be explained by the log P – $\alpha$ correlation.

**Conclusion**

The $V/V_m$ test applied to a sample of compact steep-spectrum sources ($\alpha > 0.5$) indicates that these sources evolve at a rate intermediate between that for compact flat-spectrum sources and extended steep-spectrum sources. A division of the sample into more steep ($\alpha \geq 0.7$) and less steep ($\alpha < 0.7$) sources shows that the former class evolves about as much as the extended steep-spectrum sources and the latter about as much as the compact flat-spectrum sources. This means that the phenomenological division of extragalactic radio source spectra at $\alpha = 0.5$ into flat and steep may not be as useful as a division at a higher value (0.7?) of $\alpha$. The log P – $\alpha$ correlation observed in steep-spectrum sources cannot explain the difference in evolutionary behaviour. It should be stressed that these conclusions rest on results for rather small samples and should be viewed with caution.

*Acknowledgments*
We thank Vasant Kulkarni for critical comments.